 \documentclass[showpacs,amsmath,amssymb]{revtex4}

\usepackage{graphicx}
\usepackage{dcolumn}
\usepackage{bm}

\begin{document}

\title 
      {Clean measurements of the nucleon axial-vector and free-neutron magnetic form factors}

%\classification{14.20.Dh}
%\keywords{Axial form factor, electromagnetic form factors, beta decay, Inverse reaction, parity violation}%{Document processing, Class file writing, \LaTeXe{}}

\author{A. Deur}{
 % address={Thomas Jefferson National Accelerator Facility, Newport News, Virginia 23606, USA},
 % email={deurpam@jlabt.org}
}

\affiliation{
\baselineskip 2 pt
\centerline{{Thomas Jefferson National Accelerator Facility, 
Newport News, VA 23606}}
}


%\copyrightyear  {2001}

\begin{abstract}
We discuss the feasibility of a weak charged current experiment using
a low energy electron beam. A first goal is to measure the Q$^{2}$
dependence of the axial-vector form factor $g{}_{a}(Q^{2})$. It can
be measured model-independently and as robustly as for electromagnetic
form factors from typical electron scattering experiments, in contrast
to the methods used so far to measure $g{}_{a}(Q^{2})$. If $g{}_{a}(Q^{2})$
follows a dipole form, the axial mass can be extracted with a better
accuracy than the world data altogether. The most important detection
equipment would be a segmented neutron detector with good momentum
and angular resolution that is symmetric about the beam direction,
and covers a moderate angular range. A high intensity beam
(100 uA) is necessary. Beam polarization is highly desirable as it
provides a clean measurement of the backgrounds. Beam energies
between 70 and 110 MeV are ideal. This range would provide a $Q^{2}$
mapping of $g{}_{a}$ between 0.01 <$Q^{2}$< 0.04 GeV$^{2}$. 60
days of beam can yield 14 data points with a subpercent statistical
and point to point uncorrelated uncertainties on each point. Such an
experiment may also allow to measure the free-neutron magnetic form
factor $G_{M}^{n}$. The experiment employs the usual techniques of
electron-nucleon scattering and presents no special difficulty. Higher
energy extensions are possible. They could yield
measurements of $g{}_{a}(Q^{2})$ up to $Q^{2}$=3 GeV$^{2}$ and
the possibility to access other form factors, such as the almost unknown
pseudoscalar form factor $g_P$. However, the experiments become much
more challenging as soon as beam energies pass the pion production
threshold.
\end{abstract}

\date{\today}

\maketitle

\section{Motivation}

Form factors are fundamental quantities describing hadrons and provide
crucial insight into their structure. Their precise measurements are
benchmarks for the theories and phenomenologies  aiming at 
describing the hadronic and nuclear structure, such as
Lattice QCD, Chiral Perturbation Theory or QCD counting rules.

Lepton scattering off a nucleon is described by 4 form factors: The electromagnetic
form factors $G_E^N(Q^2)$ and $G_M^N(Q^2)$ (where $N$ indicates the proton 
or the neutron) and the axial-vector and  induced pseudoscalar form factors
$g_A(Q^2)$ and $g_p(Q^2)$. $g_p$ is almost unknown and 
is interpreted as arising from scattering off the meson cloud made dominantly of 
pions. $g_A$ is better known than $g_p$ but much less than $G_E^N$ and 
$G_M^N$, although it is of the same importance. In particular, precise lattice QCD 
predictions exist for $g_A$, see e.g. [1]. %[\ref{ga lattice}]. 
There is no
precise and accurate data for $g_A$ because it is measured from either neutrino 
elastic scattering (with both weak charged and neutral currents), or pion 
electroproduction. Neutrino experiments are delicate to carry out, accumulate 
statistics slowly, relies on Monte Carlo simulations and are done typically on dense 
nuclear targets such as iron, rather than free nucleons, although the most recent 
experiments were carried on 
lighter targets ($^{12}$C and $^{16}$O). Pion data need model-dependent 
corrections to be interpreted. Indeed, neutrino and pion data disagreed until 2002 
when new corrections based on baryon chiral perturbation theory seemed to solve 
the disagreement, see [2]. %[\ref{Meissner}]. 
It is important to independently check these corrections. Furthermore, new 
tensions between experimental results arose with the most recent neutrino 
experiments that found a shallower $Q^2$-dependence of $g_A(Q^2)$. 
Parametrizing it with a dipole form: $g_A(Q^2)=g_a/(1+Q^2/M_A^2)^2$, defines 
the nucleon axial mass $M_A$. Earlier experiments measured $M_A = 1.03 \pm 
0.02$ GeV while the recent K2K experiment [3] %[\ref{K2K}] 
yields  $M_A = 1.20 \pm 0.12$ from $^{16}$O and $M_A = 1.14 \pm 0.11$ from 
$^{12}$C and the MiniBooNE collaboration measured $M_A =1.35 \pm 0.17$ from 
$^{12}$C [4]. %[\ref{miniboon}].

Consequently, it is important to provide a third, more robust, way to measure 
$g_A(Q^2)$. To do so, we propose to measure the weak charged current reaction
 $e+p\rightarrow\nu+n$ using a $\sim$100
MeV electron beam. Some of the material presented here is from [5], % {[}\ref{neptune LOI pac 25}{]}
which discusses a similar (but more difficult) experiment with GeV electron beams.

Another motivation for this measurement is to obtain the free-neutron form factors. 
Neutron structure information is so far extracted from nuclear targets (D, $^3$He), 
which involves nuclear corrections. It is obviously desirable to obtain the 
information from a free neutron. Related interests in this program, such as 
measuring $g_A(Q^2)$ in $^3$He,  investigating second class currents or 
obtaining  the ratio of the axial to vector coupling constants in a novel way,  are 
discussed in [6]. % [\ref{dutta}].
Similar programs had also been discussed in [7].%[\ref{CEBAF}].

\section{Experiment}

Although the elastic reaction $e+p\rightarrow\nu+n$ has been considered for a 
long time, no experiment has yet been done due to several difficulties. One is that 
there are only neutral particles in the final state. We will not consider detecting the 
neutrino but only the recoiling neutron. Although detection of neutrons is routinely 
done, it is difficult to determine their kinematics to a level at which the elastic 
reaction can be cleanly selected. Furthermore, a weak cross section is typically 
$10^{-11}$ times smaller than its electromagnetic (EM) counterpart. Consequently, 
the weak reaction is buried deep under the EM background. One well established 
solution to this problem is to measure the interference term between the EM and 
the weak reactions using the resulting small (typically ppm) single spin asymmetry 
(parity violating experiments, see e.g. the reviews [8]. %[\ref{PV}]).  
However, this technique only allows us to study reactions with the same final 
states as the EM reactions, that is only with the neutral weak current. Another 
strategy to reduce the background to a similar level as for the PV technique, but 
allowing to access weak charged current, is to select the backward reaction. There,  
 the undetected neutrinos recoil at large angles and the neutrons are detected at 
small angles. The Weak/EM cross section ratio is enhanced to about 
$6 \times 10^{-6}$ for a 100 MeV beam and a lepton scattering angle of 150$^o$ 
(corresponding to a nucleon recoil angle of 14$^o$. The small backward 
cross-sections require luminosities of $10^{39}$ to $10^{40}$ cm$^{-2}$s$^{-1}$. 
The beam energy should be below $\sim150$ MeV to avoid the pion 
production since this  can produce a neutron in the final state with a proton in the 
initial  one. The beam must be polarized to reduce the EM background and pulsed 
to measure the neutron energy with TOF technique and to avoid the prompt EM 
background (photon flash).  We list below the main experimental components:

\noindent \textbf{Beam:}  We assume a 63 MHz beam structure (already used with 
the Jefferson Lab CEBAF beam). The average current is assumed to be 100 $\mu 
A$, leading to a peak current of  about 800 $\mu A$. The beam polarization is the 
main ingredient to cleanly subtract the EM background: The charged current 
asymmetry is 100\% and the elastic EM one is 0. Thus, helicity minus beam pulses 
allow for both the weak and EM reactions, while helicity plus pulses allow for the 
EM reaction only. Then, subtracting events from minus helicity pulses to events 
from helicity plus pulses cleanly yields the weak reaction. However, the larger the 
EM background, the longer the experiment needs to be so that the background 
statistical fluctuations are small compared to the expected precision of the 
experiment. It is thus necessary to have other means to reduce the EM background before subtracting it from the weak signal.

\noindent \textbf{Sweeping magnet:} A sweeping magnet is needed to sweep the 
protons away from the neutron detector acceptance. At the low energies 
considered, a simple warm magnet is adequate. It will also sweep away the 
electrons. Those could be disposed of  with the TOF cuts (prompt EM 
background) but at the large luminosities considered the single electron rate would
be too large for the DAQ to handle.

\noindent \textbf{Backward detector:} A high detection efficiency backward detector 
is necessary to veto out EM reactions that produced a neutron. Events with a
recoiling electron detected in coincidence with neutron are flagged (there is no 
need to cut them at trigger level because of the low counting rate. In addition, such 
data are necessary to extract $g_p$ in higher energy versions of the experiment 
discussed here). We assume a 10$^{-3}$ detection inefficiency. It can be reached 
using existing calorimeters used in JLab experiments such as E1-DVCS [9] %[\ref{E1-DVCS}] 
or PrimEx [10]. %[\ref{primex}].
 The efficiency can be further improved by adding scintillators between the 
 detector and the target.
The detector must cover a 90$^o$ to $170^o$ azimuthal angle  and a 2$\pi$ 
polar angle to match the solid angle of the neutron detector.

\noindent \textbf{Neutron detector:} The neutron detector must be typically 2m 
away from the target to allow TOF measurements compatible with the beam 
pulsed structure: It allows to separate the photons from the neutrons and leave 
enough time for all relevant neutrons to reach the detector before the next photon 
flash. The detector should have a large acceptance to yield reasonable counting 
rates. We assume a 5$^o$ to $45^o$ azimuthal angle coverage and a 2$\pi$ 
polar angle coverage. (Complete polar angle coverage is also important for 
background management, as we will discuss.) Likewise, the detection must be 
efficient. The detector should be segmented to permit angular determination. This, 
associated with the TOF, can allow a selection of the elastic reaction and rejection 
of EM inelastic backgrounds. Finally, the detector must be shielded against low 
energy backgrounds and surrounded by scintillator paddles to veto any remnant of 
charged background (e.g. rescattering of charged particles after being swept, 
cosmic rays).

\noindent \textbf{Target:} The cell must be long for high luminosity and to maximize 
cell length over window thickness. However, since the reaction vertex cannot be 
determined, the target length contributes directly to the TOF uncertainty. 20 cm is a 
good compromise. Deuterium impurities must be minimized (5 ppm is available industrially 
and is enough). 

With this setup, we estimate that the EM backgrounds should be manageable: 10 
mil Be cell windows yield a neutron background from 
Be+e$^-\rightarrow$n+e$^-$+X with a noise/signal ratio of $8 \times10^3$ at worst 
(E=0.11GeV, 45$^o$). It can be reduced to unity with the backward electron 
detector rejecting the (e-,n) coincidences and decreased further by selecting the
nucleon elastic reaction. A 5 ppm level D contamination would yield a noise to 
signal ratio of about 0.2, becoming negligible after rejection from the backward 
detector. The backward detector also reduces to a negligible level the neutron 
contamination originating from elastically scattered protons undergoing charge 
exchange while crossing the material surrounding the cell, and from quasi-elastic 
EM reaction between the Al cell walls and electrons of the beam halo or Moller 
electrons produced in the cell.

\begin{figure}
\label{results}
\resizebox{24pc}{!}{\includegraphics{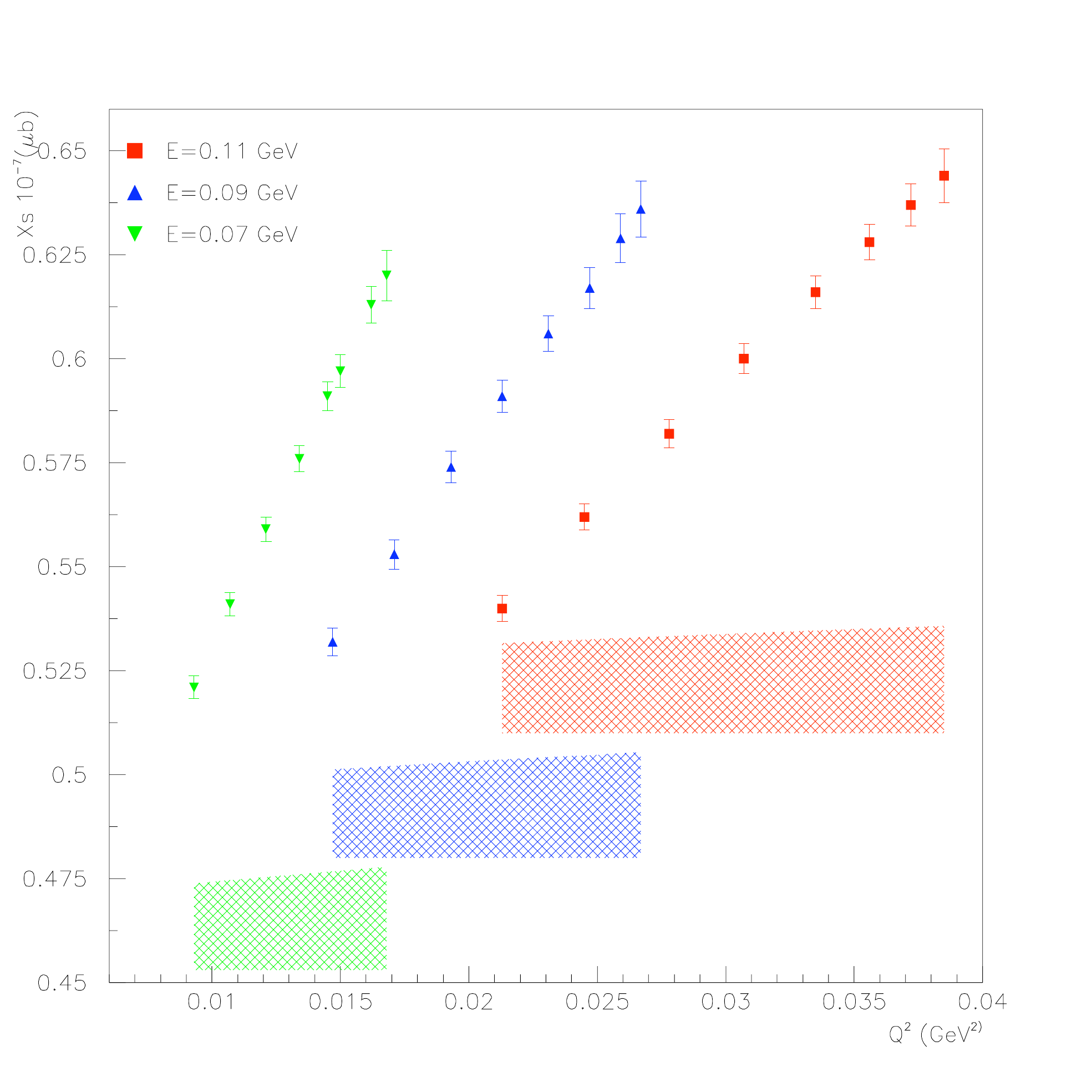}}
\caption{Expected cross-sections and uncertainties for 6 days of data taking at 110 
MeV, 7 days at 90 MeV and 17 days at 70 MeV. The error bars represent the 
statistical uncertainty and the bands the systematic ones. The uncertainties 
assume a 100\% experimental efficiency and negligible backgrounds. }
\end{figure}

In Fig. \ref{results} we show the expected cross-sections with uncertainties using 
the above experimental setup and for 6 days of running at 110 MeV, 7 days at 
90 MeV and 17 days at 70 MeV. The error bars are statistical only and the bands 
represent an assumed  4\% systematics. For the statistical errors, we took a 100\% 
experimental efficiency (neutron detector, beam polarization) and a 
signal/backgroud$\leq1$.  The typical reaction rates are a few \% of Hz for 
the weak reaction and a few $10^4$ Hz for the EM background. From such data, 
14 different  $Q^2$ points  can be extracted with an additional 10 overlapping 
points. This would provide a $Q^2$ mapping of $g_A(Q^2)$ with 
unprecedented precision. One can compare the subpercent statistical uncertainty 
of a single point to the \emph{full}  statistical precision of $\sim$13\% of the latest 
neutrino experiment [4].

In principle, at the same $Q^2$ but different angles (i.e. beam energies), $g_A$ 
and $G_M^n$ can be separated. ($G_M^p$, $G_E^p$ and $G_E^n$ must be 
taken from the world data. Generally, the contribution from the $G_E$ is small.)
However, with the energies considered here, there is not enough kinematic
lever arm for such separation. A procedure to obtain $G_M^n$ is to model
$g_a$ (e.g. with a dipole form) using our accurate mapping and assume this form
to extract $G_M^n$. The caveat of this procedure is that $G_M^n$ then depends 
on the form assumption. At higher beam energies ($\sim$1 GeV), a kinematic
separation is possible and $G_M^n$ can be obtained model independently.
However, $G_E^n$ is still out of reach because at these energies, the contribution
from the electric form factors to the cross section becomes negligible.

\section{Higher energy experiment}
It is desirable to also perform the experiment at beam energies near a few GeV  
because it would provide $g_A(Q^2)$ in an unmeasured $Q^2$ domain and 
permits a model-independent separation of $g_A$ and $G_M^n$. Furthermore, 
the backward electrons detected in coincidence with neutrons can allow access 
to the almost unknown pseudo-scalar form factor, $g_P$, up to $Q^2$ of a few 
$GeV^2$. Presently, only 3 data points exist at $Q^2<0.15 GeV^2$ [11]. %[\ref{gp}]. 
 At 1 GeV, the weak/EM cross section ratio is larger  by a  factor $\sim30$ but  
new experimental difficulties arise that actually make the experiment more 
difficult. An inelastic EM background with neutrons present in the final state 
appears above the pion production threshold. This EM background has a non-zero 
single spin asymmetry which can void the clean background subtraction scheme 
using a polarized beam. However, this EM asymmetry averages out when 
integrated over the polar angle. Hence, a detector setup (neutron and recoil 
detectors) symmetric around the beam line restores the subtraction scheme. 
Another difficulty is that the neutron detector should be placed at least 20 m away 
from the target to separate the relatively fast elastic neutrons from the photons. To 
keep a large solid angle we need a larger (more expensive) neutron detector. 
However, it also implies a better angular resolution and it minimizes the 
contribution of the target length to the TOF uncertainty. Nevertheless, the 
experiment is clearly challenging  at high energy. It is thus necessary to 
gain experience from the low energy experiment before embarking on the 
higher energy one. 
\section{conclusion}

We discussed the feasibility of a pioneering weak charged current experiment. It 
allows to measure the axial-vector form factor $g_A(Q^2)$ with precision and 
accuracy typical of nucleon-electron scattering. It also allows the unique 
opportunity to access the free neutron magnetic form factor $G_M^n(Q^2)$. The 
experiment appears feasible without requiring any new  technology. Within two 
months of running (60\% detection efficiency, 85\% beam polarization, 
signal/backgroud$\leq1$) at a luminosity of $6 \times 
10^{38}$ cm$^{-2}$s$^{-1}$, it can provide a 14-points $Q^2$-mapping of $g_A$ 
from 0.009 to 0.039 GeV$^2$ with a sub-percent statistical uncertainty on 
each point. The systematic uncertainty is expected to be 4\%. From this mapping, 
if the dipole behavior of $g_A(Q^2)$ is confirmed, the axial mass can be extracted 
with negligible statistical uncertainty. This experiment would also be a stepping 
stone to more challenging higher energy experiments using charged current with 
electron beams. These would allow $g_A$ to be mapped in an unmeasured 
$Q^2$ range, to extract model-independently $G_M^n$ and to measure the 
elusive induced pseudoscalar form factor $g_p$. Such experiments at low and 
higher energy open new possibilities for nucleon structure study and searches 
beyond the standard model.

%\begin{theacknowledgments}
\textbf{Acknowledgments}
The work in [5] %[\ref{neptune}]
 was done in collaboration with D. Lhuillier.  We thank  D. Dutta for useful 
 discussions. Jefferson Science Associates operates the Thomas Jefferson 
 National Accelerator Facility under DOE contract DE-AC05-06OR23177. 
%\end{theacknowledgments}

\end{document}